\documentstyle[referee]{mn}

\def\lsim{\, \lower2truept\hbox{${<
\atop\hbox{\raise4truept\hbox{$\sim$}}}$}\,}
\def\gsim{\, \lower2truept\hbox{${>
\atop\hbox{\raise4truept\hbox{$\sim$}}}$}\,}

\def\r{\tilde{r}}
\def\rz{\tilde{r}_0}
\def\rs{\tilde{r}_s}
\def\R{\tilde{R}}
\def\rv{\tilde{r}_{vir}}
\def\Ra{\tilde{R}_A}

\input psfig.tex

\title[Dark Matter in Elliptical Galaxies]{The Fundamental Plane of Ellipticals: \\ 
I. The Dark Matter Connection}

\author[Borriello, Salucci \& Danese] {Annamaria Borriello$^1$, 
Paolo Salucci$^1$ \& Luigi Danese$^1$ \\          
(1)  International School for Advanced  Studies SISSA/ISAS, Trieste - Italy}

\date{Accepted ... ; Received ... ; in original ...}

\begin{document}

\maketitle

\begin{abstract}

We show that the small scatter around the Fundamental Plane (FP) of
massive elliptical galaxies can be used to derive important properties of
their dark  and luminous matter.
The {\it central velocity dispersion}  $\sigma_0$, appearing  in
(e.g.) the  Fundamental Plane, is  linked to photometric, dynamical  and 
geometrical properties of (luminous and dark) matter.
We find that,  inside the effective radius $R_e$, the matter traced by the 
light must largely dominate over the dark matter (DM),
in order to keep the ellipticals close enough to the FP. This recalls
analogous findings for spiral galaxies.

In particular we also find  that cuspy DM distributions, as predicted
by numerical simulations in $\Lambda$CDM  cosmology,  are unable to
explain the very existence of the FP; in fact, according to this theory, the
structural properties of dark and luminous matter are so interwoven that a curved
surface is predicted in the log--space ($\sigma_0$, $R_e$, $L$), rather than a
plane.  In order to agree with the FP, CDM halos must have concentrations parameters  
in the range of $5-9$ (i.e.  values  significantly lower than the 
current predictions).

Assuming a more heuristic approach and allowing for cored DM halos,
we find  that the small  intrinsic scatter of the  FP yields to
{\it i)} an average value for the dark--to--light--traced mass ratio
inside the  length--scale of light $R_e$ of about $0.3$, {\it ii)}
a mass--to--light ratio of the matter traced by the light  increasing
with spheroid luminosity: $M_{sph}/L_r \propto L_r^{0.2}$ in Gunn--$r$ band,
with a value of $ 5.3$ at $L_{\ast r} \equiv 2.7 \times 10^{10} L_{r \odot}$.

\end{abstract}

\begin{keywords}
Cosmology: dark matter halos -- Galaxies: ellipticals
\end{keywords}

\section{Introduction}     

In the hierarchical scenario, dark matter (DM) halos have driven, from a
variety of initial conditions,  a dissipative infall of baryons and formed the
galactic systems we observe today (White and Rees, 1978). Thus, we  expect 
that DM halos exist within and surrounding any galaxy, regardless 
of its luminosity and morphological type. This prediction had overwhelming
confirms for disk galaxies, due to the existence of good 
dynamical tracers and  their intrinsic simple geometry (see Persic and Salucci, 1997). 
Elliptical galaxies (E's), however, are much  more complicated objects, due to 
their  3--dimensional shape,  stellar orbital structure and velocity dispersion anisotropy. 
These factors  have made ambiguous the interpretation of observational data.

A number of different mass tracers have been used to probe 
the gravitational potential in tenth of E's and  derive their mass
distribution: integrated stellar absorption spectra,  X--ray emission from hot gas,
rotating gas disks, motions of globular  clusters or satellite galaxies and, in last years, 
weak gravitational lensing. As result,  the presence of dark matter in E's, especially  
in the external regions ($\gsim 10$ kpc), is proven (e.g. Loewenstein and White, 1999).   
On the other hand,  a kinematical  modeling of the inner regions (i.e. within the
half--luminosity ``effective" radius $R_e$), 
has been performed for only a  small number of ellipticals
(e.g. van der Marel, 1991; Saglia et al. 1992, 1993; Bertin et al., 1994;
Kronawitter et al., 2000; Gerhard et al., 2001);
the results  point to a  tendency for moderate
dark matter amounts inside $R_e$.

Since its discovery, the ``Fundamental Plane"  (Djorgovski \& Davis, 1987; Dressler et al., 1987)
has been  one of the main tools to investigate E's properties:
effective radius $R_e$, central velocity  dispersion  $\sigma_0$ and mean
effective surface brightness $I_e$ of spheroidal galaxies are linearly
related in the logarithmic space and  galaxies closely cluster  on a plane,
with a surprisingly low orthogonal scatter.
To explain these linear relations between photometric and
dynamical  quantities in log--space,
most  studies on the Fundamental Plane (FP) have considered
models in which the mass is distributed parallel to light. However, in
presence of non--baryonic dark matter, this hypothesis is an obvious
oversimplification and, at least, unjustified. Indeed, this would {\it a
priori} require either:  {\it i)}  dark and luminous component
are distributed according the same profile, thus revealing a similarity of
properties and behavior which seems very unlikely  or {\it ii)} the dark matter
component is always negligible with respect to the luminous matter.

Within the above framework, in this paper  we address
the following issues:
\begin{itemize}
\item  to derive the relation between the central velocity dispersion
$\sigma_0$ and the mass distribution parameters, including the effect of
a dark matter halo.  In particular, we assume a spherical model  with an
isotropic luminous component and a dark halo, more diffuse than the spheroid,
\item to reproduce the observed Fundamental Plane and, therefore, to constrain
the mass distribution in E's,
\item   to discuss the results in the light of Cold Dark Matter predictions.
\end{itemize}

Considering elliptical galaxies as two--components systems,
complementary strategies are possible.
One chooses a distribution function for both components and then
imposes  specific constraints from the observations. The other includes  the
ordinary stellar component  (or, better, any traced by light (TBL) mass component) 
in a frozen spherical halo. The former approach is helpful in exploring  the
self--consistency of the dynamical configuration (e.g. Ciotti, 1999).
The latter, we will adopt   in this paper, has the advantage of providing a
simpler connection  between observational quantities and the  parameters of the
mass model.

The outline of this paper is the following:  in \S2 we describe
two--components models, whose mass distributions are shown in  \S3.
In \S4  we derive and  discuss the velocity dispersion (line--of--sight profile
and central value) predicted by the mass models we consider. In \S5,
we introduce  the data,  fit the models to the Fundamental Plane and discuss the results.
Finally, conclusions are presented in \S6.
Throughout the following work, we assume, where needed,
a flat $\Lambda$CDM Universe, with $\Omega_m=0.3$, $\Omega_{\Lambda}=0.7$,
$h=0.7$ and $\sigma_8 =1.0$.

\section{The velocity dispersion in the 2-components mass models} 

The observed velocity dispersion and, in particular, 
the {\it central} velocity dispersion $\sigma_0$ is a fundamental dynamical
property related to the gravitational potential of both the dark halo
and the TBL  component (sometime,  we call the  ``luminous stellar spheroid"  by 
this way, to recall the possible existence of a number of BH's and/or an amount of 
non--baryonic DM, perfectly mixed with the luminous stellar matter). 
Let us start by assuming a spherical and
non--rotating stellar system, with the stellar velocity dispersion which is
the same  in all directions perpendicular to a given radial vector.
If $\sigma_r^2(r)$ denotes the stellar velocity
dispersion along the radial vector and $\sigma_{\theta}^2(r)$
the dispersion in the perpendicular directions, the Jeans hydrodynamic
equation  for the mass density traced by the light $\rho_{sph}(r)$
in the radial direction reads (Binney and Tremaine, 1987):
\begin{equation}
\frac{d \rho_{sph}(r) \sigma_r^2(r)}{dr} + \frac { 2\ \beta(r)
\rho_{sph}(r) \sigma^2_r(r)}{r} = -\frac{G M(r)}{r^2} \rho_{sph}(r)
\label{jeans}
\end{equation}

\noindent with the boundary condition $\rho_{sph}(r) \sigma^2_r(r) \rightarrow 0$
for $r \rightarrow \infty$.
In eq.(\ref{jeans}), the parameter
$\beta(r) \equiv 1-\sigma_{\theta}^2(r)/\sigma_r^2(r)$
describes the anisotropy degree of the
velocity dispersion at each point, with $\beta=1, 0, -\infty$
for completely radial, isotropic and circular orbit distributions,
respectively.

Dynamical analysis of
ellipticals  exclude  substantial amount of tangential anisotropy and
find  $\beta(\lsim R_e) \simeq 0.1-0.2$  (e.g. Matthias and
Gerhard, 1999; Gerhard et al., 2001; Koopmans and Treu, 2002)
for objects of different luminosities.
For reasons of simplicity, then,  in calculating the central velocity dispersion to be used   for
statistical studies over a large sample of galaxies,  we assume an
isotropic velocity dispersion tensor:  $\beta=0$.

Eq.(\ref{jeans})  connects the spatial velocity
dispersion of the component traced by the light to its density profile and
to the {\em total} matter distribution  $M(r)=M_{sph}(r)+M_h(r)$.
Under the hypothesis of isotropy, the above equation assumes the well--known
integral form, in which we single out the halo term:

\begin{equation}
\sigma_r^2(r)  =  \frac{G}{\rho_{sph}(r)} \int_r^{\infty} \frac{\rho_{sph}(r') M(r')}{r'^2}\ d r'
\equiv   \sigma_{r;sph}^2(r) + \sigma_{r;h}^2(r)
\label{sigmar}
\end{equation}

As external observers of galaxies, we measure only projected
quantities. Let $R$ be the projected  radius and  $\Sigma(R)$ the
surface stellar mass density. As usual, we take into account that
mass density and the spatial velocity dispersion are related to
the surface mass density $\Sigma(R)$ and to the projected velocity
dispersion $\sigma_P(R)$ by the two Abel integral equations for the
quantity $\rho_{sph}$ and $\rho_{sph}\sigma_r^2$. Then, a
second step, consisting in  a further integration along
the line of sight, allows us to obtain the (observed) velocity dispersion
profile $\sigma_P(r)$:

\begin{equation}
\sigma_P^2(R)  =  \frac{2}{\Sigma(R)}\ \int_R^\infty\ \frac{\rho_{sph}(r)\ \sigma_r^2(r)\ 
r}{\sqrt{r^2-R^2}}\ d r  \equiv \sigma_{P;sph}^2(R)+ \sigma_{P;h}^2(R)  
\label{sigmap}
\end{equation}

\noindent where $\Sigma(R)=\int_R^\infty\ [2\ r\ \rho_{sph}(r)/(r^2-R^2)^{1/2}]\ d r$.

As spectro--photometric observations are performed through an aperture,
let us define  $\sigma_A(R_A)$  as the luminosity--weighted 
average of $\sigma_P$ within a circular aperture of radius $R_A$:

\begin{equation}
\sigma_A^2(R_A)   =   \frac{2 \pi}{L(R_A)}\ \int_0^{R_A}\ \sigma_P^2(R)\ I(R)\ R\ dR  
\equiv   \sigma_{A;sph}^2(R_A) + \sigma_{A;h}^2(R_A)
\label{sigmaap}
\end{equation}

\noindent where $I(R)$ is the surface brightness profile  $I(R)=\Sigma(R)/\Upsilon$
(assuming the stellar mass--to--light ratio $\Upsilon$ constant  with radius)
and $L(R_A) = 2\ \pi\ \int_0^{R_A} I(R)\ R\ dR$ is the aperture luminosity. 

The dynamical quantity in the Fundamental Plane is    
the ``central" velocity dispersion $\sigma_0$, which, observationally,
corresponds to {\it the projected velocity  dispersion 
luminosity--weigthed  within the aperture of the observations}.
The velocity dispersion data are to be brought  to a  common
system, independent of the telescope and galaxy distance: this is done by
correcting  them  to the same  aperture of $R_e/8$  (J{\o}rgensen et al.,
1996), which is typical of  measurements of nearby galaxies. Therefore, we
compare model and observations by calculating  $\sigma_0$
as the  luminosity--weighted $\sigma_P(R)$
within $R_A = 1/8\ R_e$.

The resulting velocity dispersion profiles, $\sigma_r(r)$,
$\sigma_P(R)$ and $\sigma_A(R_A)$, can be all expressed
as the sum of two terms  (eqs. \ref{sigmar}, \ref{sigmap}
and \ref{sigmaap}): the first one is due to the self--gravity of the
spheroid traced by the star light (labelled by $sph$); the second one 
(labelled by $h$) is due to the effect of the luminous--dark matter gravitational
interaction and, therefore, it charges relevance
according the characteristics of the  DM distribution.

\section{The mass distribution}

\subsection{The  distribution of the mass  traced by light  \label{stdist}}

We describe the  component traced by the star light by means of a
Hernquist (1990) spherical density distribution, that  is a good
approximation to the de Vaucouleurs  $R^{1/4}$ law (de Vaucouleurs, 1948)
when projected and, at the same time,
allows analytical calculations:
\begin{equation}
\rho_{sph}(r)=\frac{M_{sph}}{2\pi} \frac{k\ R_e}{r\ (r+k\ R_e)^3}
\label{rhostar}
\end{equation}
where $M_{sph}$ is the total mass traced by the star light and  
$k \simeq 0.55$. 
The mass profile  derived from eq.({\ref{rhostar}) is:
\begin{equation}
M_{sph}(r)=M_{sph} \frac{(r/R_e)^2}{(r/R_e+k)^2}
\label{Mstar}
\end{equation}
The Hernquist functional form, of course, cannot reproduce the fine features
of the surface brightness profile (e.g. boxy isophotes, small variations in slope);
neverthless, it is  sufficient  for our aims, since we will just consider large
scale properties in the mass distribution of objects belonging to a large
E's sample.

\subsection{The DM distribution: $\Lambda$CDM halos}

N--body simulations of hierarchical
collapse and merging of CDM halos have shown that gravity,
starting from scale--free initial conditions, produces an universal density profile
that, for $r\rightarrow\ 0$, varies with radius as $r^{-\alpha}$, with
$\alpha \sim 1-2$ (e.g. Navarro, Frenk \& White, 1997, hereafter NFW;
Fukushige and Makino, 1997; Moore et al., 1998; Ghigna et al., 2001),
weakly dependent on the cosmological
model.  We adopt  the well--known  NFW  halo density profile:
\begin{equation}
\rho_{NFW}(r)=\frac{\rho_s}{(r/r_s)(1+r/r_s)^2}
\label{rhoNFW}
\end{equation}
where $r_s$ is the inner characteristic
length--scale, corresponding  to the radius where the  logarithmic
slope of the profile  is $-2$.
It results convenient
to write the NFW mass profile as:
\begin{equation}
M_{NFW}(r)= M_{vir}\ \frac{A(r, r_s)}{A(c, r_s/R_e)}
\label{MNFW}
\end{equation}
where $A(x, y) \equiv \ln (1+x/y) - (1+y/x)^{-1}$
for any pair of variables $(x, y)$.
The concentration parameter is defined as $c \equiv r_{vir}/r_s$;
$r_{vir}$ and $M_{vir}$ are, respectively,
the halo virial radius and mass.
The definition of the virial radius is strictly  within the framework of
the standard  dissipationless spherical collapse model (SCM);
however, also in more realistic  hierarchical models, it provides a
measure  of the boundary of virialized  region of halos
(Cole and Lacey, 1996).

Considering a dark halo at redshift $z$, the virialized region
is the sphere within  which the  mean density is $\Delta_{vir}(z)$  times the background
universal density  at that redshift ($\rho_{bkg}=\rho_c\ (1+z)^3$,
with $\rho_c$ the critical density for closure at $z=0$).
The virial mass is defined as:
$M_{vir} \equiv \frac{4}{3}\ \pi \ \Delta_{vir}(z)\ \rho_{bkg}\ r_{vir}^3$,
with the virial overdensity $\Delta_{vir}$ being a function both of  the
cosmological model  and  the redshift: for the  family
of flat cosmologies ($\Omega_{\rm m}+\Omega_{\Lambda}=1$), it can be
approximated by (Bryan and Norman, 1998):
$ \Delta_{vir}(z) \simeq 18 \pi^2 + 82 (\Omega(z)-1) -
39 (\Omega(z)-1)^2/\Omega(z)$.
From the above equations, we derive:
\begin{equation}
r_{vir} = 0.142\ \Delta_{vir}(z)^{-1/3} (1+z)^{-1} \left
(\frac{M_{vir}}{M_{\odot}} \right )^{1/3}\ h^{-2/3}\ {\rm kpc}
\label{Rvir}
\end{equation}

A fundamental result of  CDM  theory  is that  the
halo concentration  well correlates with the  virial mass:
low--mass halos are denser and more concentrated than high--mass halos
(Bullock et al., 2001; Cheng and Wu, 2001; Wechsler et al., 2002)
in that, in average, they collapsed when the Universe was denser.
Numerical experiments by  Wechsler et al.
(2002)  for a population of halos identified at
$z=0$ show that:
 
\begin{equation}
c(M_{vir}) \simeq c_{11}\ \left ( \frac{M_{vir}}{10^{11}\ M_{\odot}} 
\right )^{-0.13},
\end{equation}
with $c_{11}\simeq 20.8$. The  Poisson error for galactic  halos in the 
mass range $M_{vir} \lsim 10^{11-13} M_{\odot}$ is less than 10$\%$ and, 
virtually, no halo is found with $c < 12$.

This is the  first mass model we consider here: it 
is composed by a stellar bulge with a Hernquist  profile embedded in a spherical
dark NFW halo (hereafter H+NFW model).   We neglect
the effects of a possible adiabatic coupling between baryons and dark matter
because it  has the effect of increasing  the DM density  inside $R_e$
by adiabatic compression and, then, of worsening the fit to the data (see
$\S$\ref{NFWres}).
The ``total" dark--to--TBL mass ratio, defined as  $\Gamma_{vir}
\equiv M_{vir}/M_{sph}$ is  a crucial parameter of the mass model.
Both BBN predictions about the primordial DM--to--baryons ratio and 
CMB anisotropy observations point at a lower limit for $\Gamma_{vir}$ 
of $\simeq 8$.

\subsection{The DM distribution: cored halos}

In last years, studies of high resolution rotation curves of spiral
and dwarf galaxies casted doubts on the presence of the
central cusps predicted by cosmological simulations of DM halos.
Actually, the observational results suggest
that  dark halos are  more  diffuse than the luminous component
and their  densities flatten at small radii, whereas the stellar
distribution peaks towards the centre  (e.g. Moore, 1994; Flores and Primack, 1994;
Burkert, 1995;  de Battista and Sellwood, 1998; Salucci and Burkert, 2000;
de Block, McGaugh and Rubin, 2001; Borriello and Salucci, 2001).

An useful analytic form for halos with soft cores has been proposed
by Burkert  (1995) for dwarf galaxies and, then, was  extended to the
whole family of spirals  by Salucci and Burkert (2000):
\begin{equation}
\rho_B(r)=\frac{\rho_0}{(1+r/r_0)[1+(r/r_0)^2]}
\label{rhoB}
\end{equation}
This  profile is characterized by  a density--core of extension $r_0$ and
value $\rho_0$, while it  resembles the NFW profile at large radii.
From eq.(\ref{rhoB}) the mass profile  reads:
\begin{equation}
M_B(r)= M_e\  \frac{B(r, r_0)}{B(1, r_0/R_e)}
\label{MB}
\end{equation}
where, for any pair of variables $(x , y)$,
$B(x,y)\equiv -2 \arctan (x / y)  + 2 \ln (1+x / y) + \ln [1+(x/y)^2]$  and
$M_e$ is the dark mass within $R_e$.
Then, in analogy of spiral galaxies, we propose a  mass model
consisting of a Hernquist  bulge plus a Burkert dark halo (hereafter H+B
model).  This is the second mass model we investigate in this paper; 
let us notice that, in this case, the  halo mass distribution is characterized  by two free
parameters:  the dark--to--TBL  mass ratio within the effective radius
$\Gamma_e\equiv M_{DM}/M_{sph}\mid_{R_e}$  and the halo core radius in
units of  the effective radius $r_0/R_e$. We assume  $r_0 \gsim R_e$ to ensure a
constant DM density  in the region where the stars reside:  otherwise the  model
would essentially coincide with the NFW one.  

It is worth noticing that  the Burkert profile  is purely heuristic and  
featureless in the optical regions of galaxies; 
as a consequence, unless data at large radii are
available,  it is not possible to  determine the halo virial radius and
the total galaxy mass.

\section{Mass--velocity dispersions relations for cuspy and cored models}

We compute the velocity dispersion profiles, 
including the  effect of  a spherical dark halo, for both 
H+NFW and H+B cases.   We resolve eq.(\ref{sigmar}),
eq.(\ref{sigmap})  and eq.(\ref{sigmaap}) by  assuming  the  density/mass 
profiles of eq.(\ref{rhostar}), eq.(\ref{Mstar}) and eq.(\ref{MNFW}) in H+NFW
case and  of eq.(\ref{rhostar}), eq.(\ref{Mstar}) and eq.(\ref{MB}) in
the H+B  one. The  detailed  calculus in given in  Appendix.

\begin{figure*}
\vspace{-1.7truecm}
\centerline{\ \hspace{-2truecm} \psfig{file=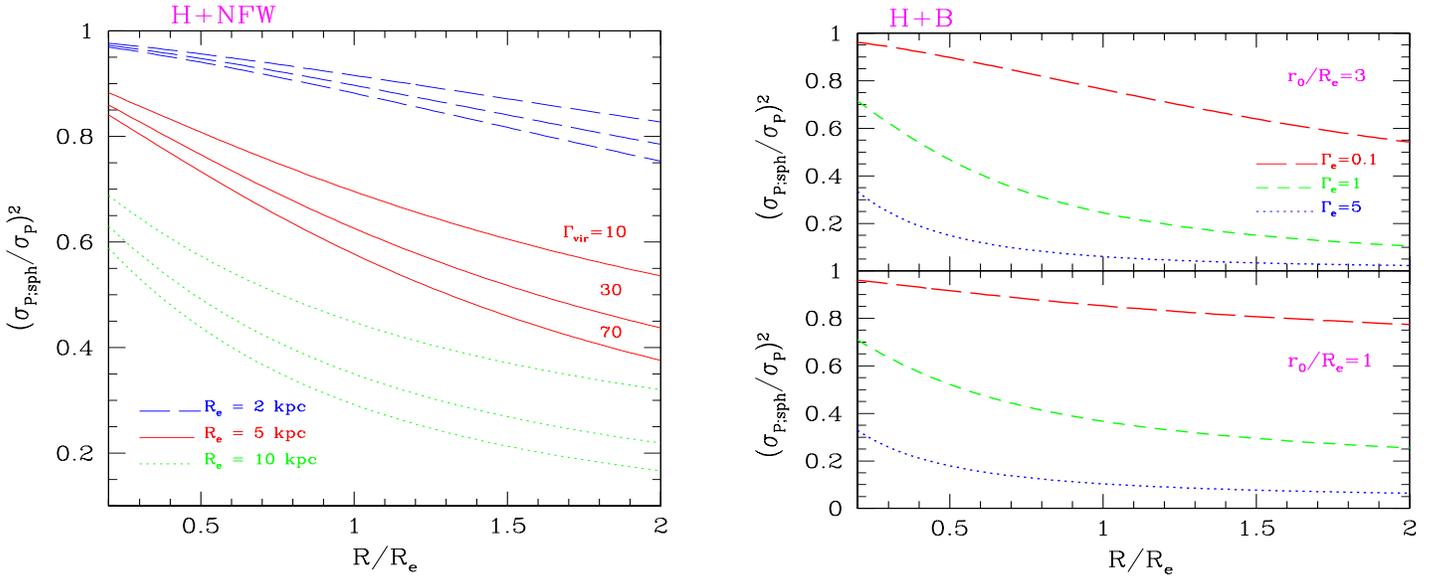,width=230mm,height=280mm}}
\vspace{-17truecm}
\caption{Line--of--sight velocity dispersion due to the   spheroid
 (in units of  {\it total}  velocity  dispersion)  as a function
of radius for: {\it left)} H+NFW case (different  values of
$R_e$ and $\Gamma_{vir}$ are indicated); {\it right)}  H+B case 
(different values of $r_0/R_e$ and $\Gamma_e$ are indicated).}
\label{sigPNB}
\end{figure*}

In Fig.\ref{sigPNB} ({\it left})  we show, for H+NFW models, the radial profile 
of the quantity $\sigma^2_{P;sph}/\sigma^2_P$, the line--of--sight  
velocity dispersion due to the TBL component, in units of  the total l.o.s. 
velocity dispersion.
We take $M_{sph}=2 \times 10^{11} M_{\odot}$; however,
the mass dependence  is very weak and the curves in Fig.\ref{sigPNB} ({\it left}) are well
representative of those with stellar  masses  in the range
$\sim 5 \times10^{9}- 10^{12} M_{\odot}$.
We consider different  plausible  values for the  total dark--to--TBL mass
ratio $\Gamma_{vir}$ and $R_e$. Let us notice that,
once we  fix the  total halo mass $M_{vir} \equiv \Gamma_{vir}\cdot
M_{sph}$, the halo characteristic radius $r_s$ is completely determined
via the $c-M_{vir}$ relationship: therefore, different  curves in
Fig.\ref{sigPNB} ({\it left})  correspond  to $2 \lsim r_s/R_e \lsim 30$.

We realize  that the contribution of a CDM halo to the velocity
dispersion can be  large ($\sim 50\%$) even at small radii $R \lsim
R_e/2$, in galaxies with large effective radii  and/or small values of
$r_s/R_e$, indipendently   of  the value of $\Gamma_{vir}$.

\begin{figure*}
\vspace{-6truecm}
\centerline{\ \hspace{-1.3truecm}
\psfig{file=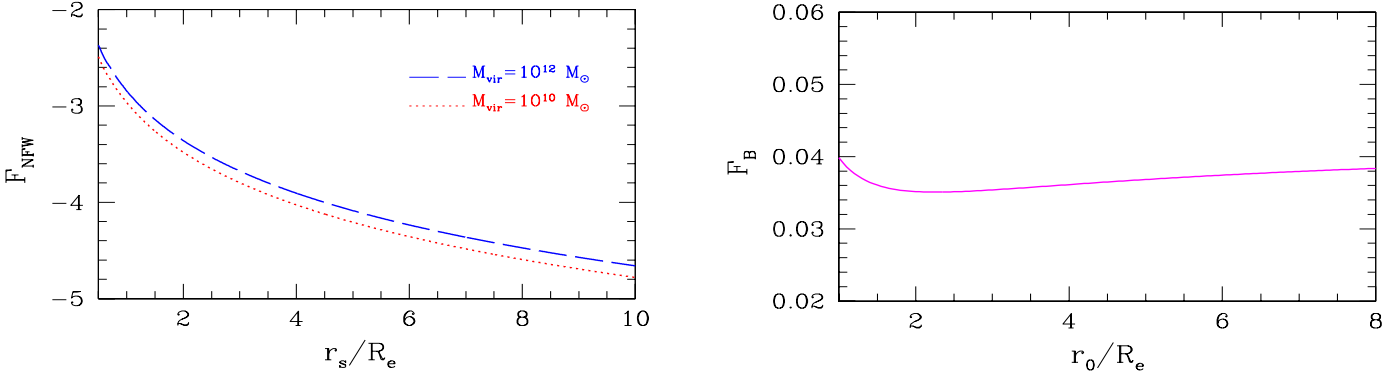,width=205mm,height=260mm}}
\vspace{-14.3truecm}
\caption{The functions $F_{\rm NFW}$ and $F_{\rm B}$ entering the expression of
$\sigma_0$, for H+NFW  model ({\it left}) and H+B model ({\it right}).}
\label{FNFB}
\end{figure*}

In Fig.\ref{sigPNB} ({\it right}) we plot the TBL--to--total  
l.o.s. velocity dispersion  ratio  for H+B models, for different 
values of the parameters $r_0/R_e$ and $\Gamma_e$.  Notice that the profiles just depend
on these parameters and it is not necessary to assume specific values
both for   $M_{sph}$ or $R_e$. The main consequence of the smooth
halo profile in the H+B model is that the halo contribution to $\sigma_P(R)$
is low at small $R$, even for models with a relevant
amount of DM  in the central region  (e.g. $\sigma^2_{P;sph}/\sigma^2_P
\sim 80\%$ at $R_e/3$ when $\Gamma_e=1$). The velocity
dispersion in the central regions is more directly connected to
the properties of the TBL  mass distribution.

The relationship of the spheroid mass with central velocity dispersion
$\sigma_0$ and the effective radius $R_e$ depends on the DM mass
distribution.
Recalling that  $\sigma_0 \equiv \sigma_A(R_e/8)$,  we find:

\begin{equation}
\sigma_0^2= (0.174 + \Gamma_{vir}\ F_{NFW})\ \frac{G\ M_{sph}}{R_e}
\ \ \ \ \ \ \ \ \ \ \ \ \ \ \ {\rm H+NFW}
\label{sigma0NFW}
\end{equation}
\begin{equation}
\sigma_0^2= (0.174 +  \Gamma_{e}\ F_{B})\ \frac{G\ M_{sph}}{R_e}
\ \ \ \ \ \ \ \ \ \ \ \ \ \ \ \ \ \ \ \ \ {\rm H+B}
\label{sigma0B}
\end{equation}
where $F_{\rm NFW}$ and $F_{\rm B}$ are shown in Fig.\ref{FNFB}.
$F_{\rm NFW}$ depends on
$r_s(M_{vir})/R_e$ and, very weakly, on $M_{vir}\equiv \Gamma_{vir}\cdot M_{sph} $.
The term in the r.h.s. of eqs.(13) and (14) shows that, according to the values 
of the model parameters,
$\sigma_0$ could  be strongly affected  by the DM gravitational
potential, so that  the derivation of the spheroidal mass from
obsevational properties  critically depends on the actual DM mass
profile. This is shown in (Fig.\ref{sig0NB}, {\em left}), where we
plot the TBL--to--total $\sigma^2_0$, assuming $M_{sph}= 2 \times 10^{11} M_{\odot}$ and
different model parameters: the TBL contribution to $\sigma^2_0$ is
under dominant also at small $r_s/R_e$ and becomes almost negligible
for $r_s/R_e\lsim 1$ and $\Gamma_{vir} \gsim 30$.

\begin{figure*}
\vspace{-3.3truecm}
\hspace{-0.4truecm}
\centerline{\psfig{file=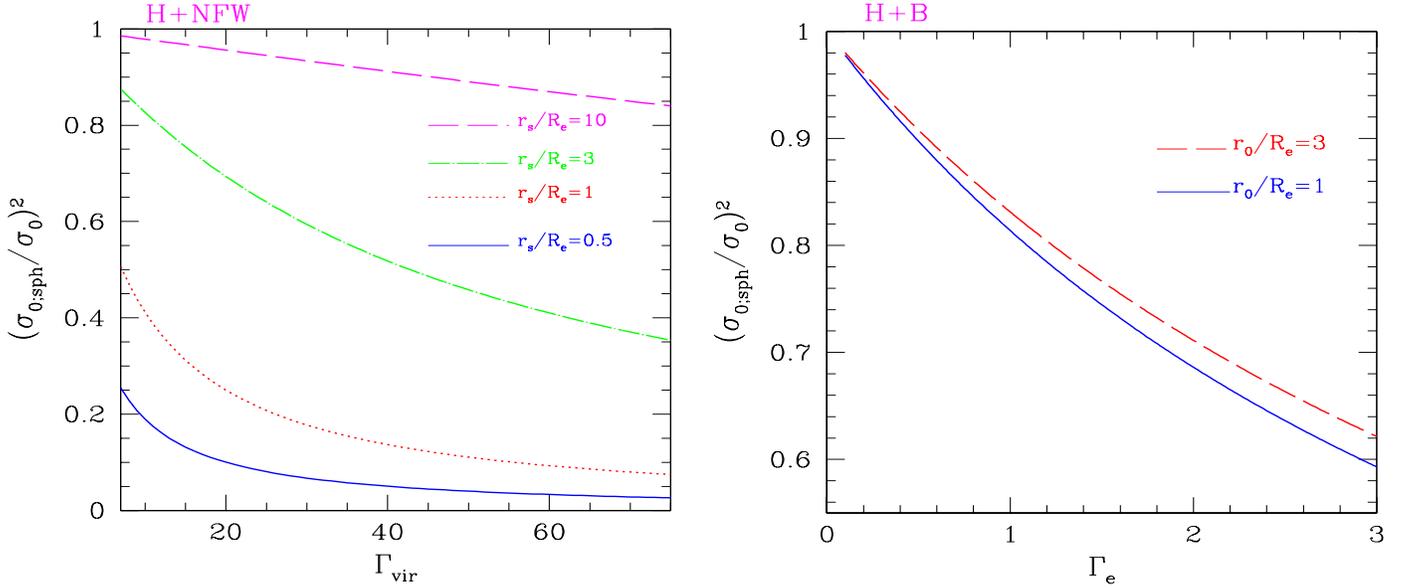,width=210mm,height=260mm}}
\vspace{-14.2truecm}
\caption{The  TBL mass component contribution to $\sigma_0$
for different models parameters: {\em left}) H+NFW predictions for
$M_{sph}= 2 \times 10^{11} M_{\odot}$, as function of
$\Gamma_{vir}$ and for different $r_s/R_e$;
{\em right})  H+B predictions, as function of the parameters $\Gamma_e$ and $r_0/R_e$.}
\label{sig0NB}
\end{figure*}

In H+B mass models, $F_{\rm B}$ weakly depends on $r_0/R_e$, so that we can assume
$F_{\rm B} \simeq 3.6 \times 10^{-2}$;
the  TBL contribution  to $\sigma^2_0$  (Fig.\ref{sig0NB}, {\em right})
remains the dominant one, even for an amount of DM
within $R_e$ comparable to the luminous one.  
Moreover, from eq.(\ref{sigma0B}) we infer that, for cored configurations (i.e. with $r_0 > R_e$),  
$\sigma_0$  is {\it weakly} dependent on the DM internal amount ($F_{\rm B}$ is of the 
order of $10^{-2}$):  
as a matter of fact, it just increases of $\sim 30\%$
when  $\Gamma_e$ varies of a factor 3. This is  a  natural consequence  of the
smoothness of the  dark matter distribution with respect to the more
concentrated distribution of the luminous spheroid.

\section{Fitting mass models to the Fundamental Plane}

\subsection{The Sample}

We build the data sample from  several works by
J{\o}rgensen, Franx \& Kjaergaard  (hereafter JFK).
They provide spectroscopy and  multicolour CCD surface photometry of E/S0
galaxies in nearby clusters. The photometric data are  from JFK (1992) and (1995a)
in Gunn--$r$, their passband with the largest quantity of data.  The
spectroscopic measurements are taken from  JFK (1995b) and references
therein. Out of the whole JFK sample, we selected
a homogeneous  subsample of 221 E/S0 galaxies in 9 clusters,
including Coma, whose properties are shown in electronic form
in Tab.1 at the URL: {\it www.sissa.it/ap/ftp}.  In particular, we rejected
spiral,  interacting, peculiar and field galaxies (due to the greater
uncertainty of their distance).
For each cluster, we adopt the distance derived in JFK (1996).
The FP {\it r.m.s.}  scatter is 0.084 in $\log R_e$ (this is equivalent to a
$\sim 17\%$ uncertainty  in galaxy  distances  that,
then, could be the main scatter contributor). Typical
measurement errors are $\Delta \log R_e=\pm 0.045$, $\Delta \log I_e=\pm
0.064$,  $\Delta \log \sigma_0=\pm 0.036$, and  $\Delta \log L_r=\pm
0.036$, maybe  large enough  to imply that the  FP is free from an  
intrinsic scatter.

The statistical distributions of  effective radius, ellipticity at $R_e$,
observed central velocity dispersion and Gunn--$r$  luminosity of the selected
galaxies are shown in Fig.\ref{hist}.
It is worth noticing that the sample galaxies are  distributed around
$L_{\ast}=2.7 \times 10^{10} L_{r \odot}$, the characteristic luminosity of the
ellipticals luminosity function in $r$--band (Blanton et al., 2001) and that
most of the objects  have little/moderate ellipticity ($<\epsilon>=
0.29 \pm 0.17$) and then, a reasonably spherical  stellar distribution.

\subsection{Forcing models to the Fundamental Plane}

\begin{figure*}
\vspace{-1.3truecm}
\centerline{\psfig{file=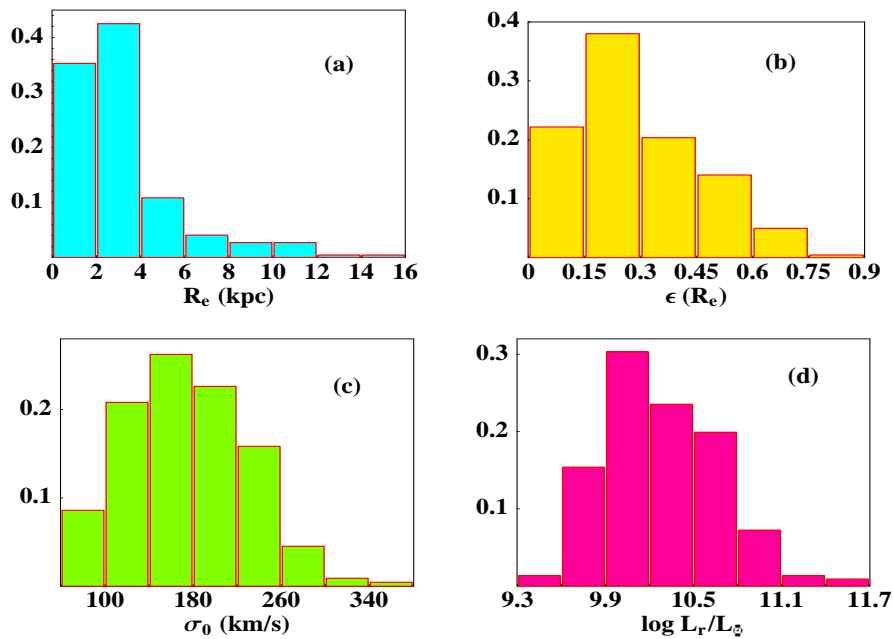,width=170mm,height=210mm}}
\vspace{-9.8truecm}
\caption{Histograms of  sample galaxies properties: (a) effective radius
$R_e$, (b) ellipticity $\epsilon$ at $R_e$, (c) central velocity dispersion
$\sigma_0$  and   (d) luminosity $L_r$ in Gunn--$r$  band .}
\label{hist}
\end{figure*}

The physical interpretation of the Fundamental Plane  assumes  the virial
theorem to be the main constraint to the structure of
ellipticals. Assuming elliptical galaxies to be 1--component and homologous systems,  
the virial theorem and the existence of the FP imply a slow  but systematic 
variation of the mass--to--light ratio $M/L$   with the luminosity,  
whose physical origin is debated.
This homology also determines a quasi--linearity of the relations connecting
the gravitational and kinematic scale parameters of the galaxies to the observables
$\sigma_0$ and  $R_e$ (see Prugniel and Simien, 1997). 
However,  according to the properties of the second, dark,
mass component, this property could be lost,  in such a way that the
gravitational and photometric scales are not anymore connected in a
simple, log--linear way.

\begin{figure*}
\vspace{-3.5truecm}
\centerline{\ \hspace{2truecm}\psfig{file=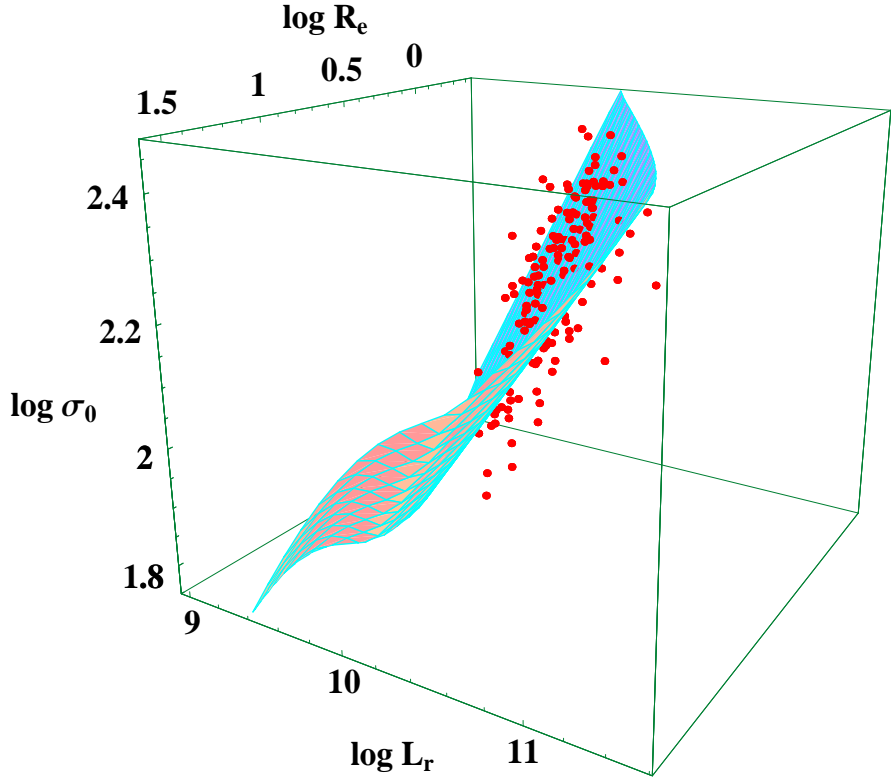,width=220mm,height=320mm} }
\vspace{-18truecm}
\caption{The surface  in the log--space ($R_e$, $L_r$, $\sigma_0$) where we expect
galaxies in the   H+NFW scenario for   $\Gamma_{vir}=30$, compared to 
sample data (red points). The units are: $R_e$ in kpc, $L_r$ in $L_{r \odot}$ and
$\sigma_0$ in km/s.} 
\label{FPN} 
\end{figure*}

We adjust  the mass models parameters to fit the observations in the
log--coordinates space:  effective radius $R_e$,  central velocity dispersion
$\sigma_0\equiv \sigma_A(R_e/8)$ and total luminosity in Gunn--$r$ band,
defined as $L = 2 \pi R_e^2 I_e$.   The effective surface
brightness $I_e$ in L$_{\odot}/$pc$^2$ is  calculated from $\mu_e$ in mag
arsec$^{-2}$:  $ \log  I_e = -0.4 \mu_e - 26.4$, for Gunn--$r$  band
(JFK, 1995a).  In fitting the log--surface $\sigma_0 (R_e,\ L_r)$
to the observations, we leave free the mass--to--light ratio  
$M_{sph}/L_r$ of the TBL component
and, respectively,  $\Gamma_{vir}$ in the  H+NFW
case  and  $\Gamma_e$ in the H+B case.
In the latter,  we assign a constant value to the parameter $r_0/R_e=2$, similar to
results for spirals (Borriello and Salucci, 2001), as the fit depends very  weakly  on it.
We characterize the  mass--to--light  ratio as
$\Upsilon_r \equiv M_{sph}/L_r = \Upsilon_{r \ast}\
( L_r/L_{\ast} )^{\alpha}$, with  $\Upsilon_{r \ast}$ and $\alpha$
free parameters.
Here, we neglect  a possible weak dependence of
$\Upsilon_r $ on $R_e$, but we will discuss this point later.

\subsection{Results and discussion}

\subsubsection{H+NFW mass model \label{NFWres}}

This model is unable to provide  a plane surface in the log--space
($\sigma_0$, $R_e$, $L_r$), for plausible values of the free parameters. 
In Fig.{\ref{FPN}}  we show the effect on the
surface by  adding  to the stellar spheroid a dark NFW component  with
the reasonable dark--to--TBL  mass ratio  of $\Gamma_{vir}=30$.
The surface curvature  prevents us from properly fitting the
data, especially in the region occupied by galaxies  with large effective
radius and low luminosity, for which the DM contribution to $\sigma_0$ is
unacceptably high.

The results of the fitting procedure are shown in Fig.\ref{CLN},
with contours  representing  $68\%$, $95\%$ and $99\%$ CL.
The best--fit model is consistent with no DM  at all
($\Gamma_{vir}= 2 \pm 4$ at 1~$\sigma$)  and values of $\Gamma_{vir}  \gsim
10$ are excluded at  $> 95\%$ CL.  Solutions  very marginally permitted
($\Gamma_{vir} \simeq 10-15$) still  require  a very high  efficiency of
collapse of baryons in stars ($\sim 90\%$).  This is at strong variance
with the inferred budget of the cosmic baryons (Salucci and Persic, 1999)
and with  current ideas of galaxy formation, for which  feedback
mechanisms (such as SN  explosions and central QSO activity)  transfer
thermal energy  into the ISM and inhibit  an efficient star formation
(e.g. Dekel and Silk, 1986;  Romano et al., 2002). They predict 
$\Gamma_{vir}$ to be significantly  higher than the cosmological  value
of $\sim 8$.

\begin{figure*}
\vspace{0.5truecm}
\centerline{\hspace{0.5truecm}\psfig{file=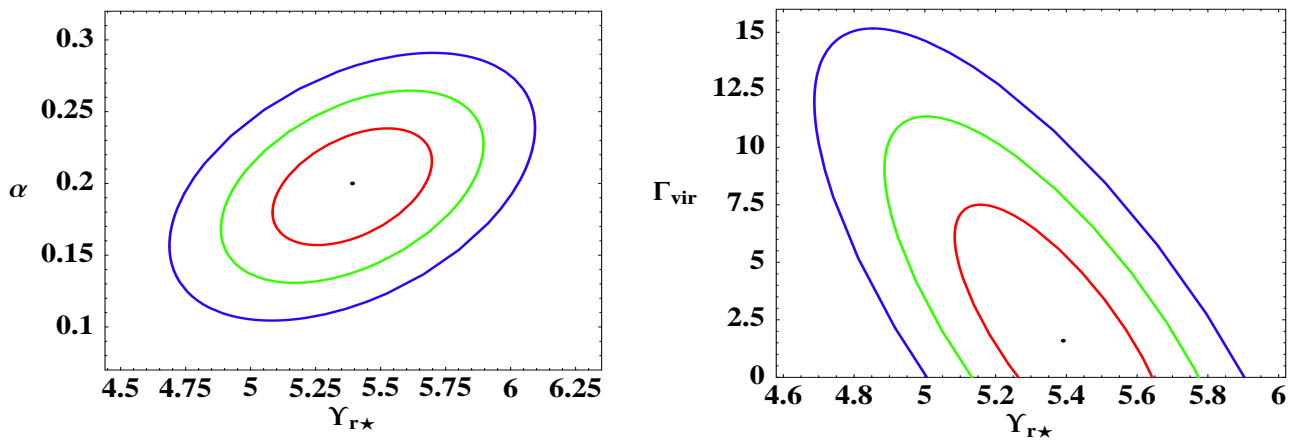,width=200mm,height=280mm}}
\vspace{-21.4truecm}
\caption{The mass model parameters, obtained by best--fitting the H+NFW 
surface in the log--space ($R_e$, $L_r$, $\sigma_0$) to the FP ($68\%$, $95\%$ and $99\%$ CL).}
\label{CLN}
\end{figure*}

The mass--to--light ratio of the TBL component in Gunn--$r$ band is found  
$\sim 5$, i.e.  well within values predicted by passive evolution of  an
old stellar population, generated with a  standard IMF (e.g. Trager et al., 2000).
As a matter of fact, this encourages us to consider the TBL mass as composed of
just stellar populations.

The curvature of the surface in the logarithmic space of the observables
is a consequence of the particular concentration--mass relation predicted
by CDM N--body simulations. 
Anyway, once we assume a value for $\Gamma_{vir}$, we can consider different 
concentration parameters (for example, by moving a fraction of DM outside $R_e$).
In order to illustrate the consequences, we have explored the realistic case
of $\Gamma_{vir}=30$.
For simplicity, we assume  $c= c_{11} \cdot (M_{vir}/10^{11} M_{\odot})^{-0.13}$, 
and we look for values of $c_{11}$ for which CDM profiles are in agreement with 
the Fundamental Plane. 
In Fig.\ref{CL30} we show the  value for  $c_{11}$  implied  by
the narrowness of the observed FP. In order to have a small  DM fraction
inside $R_e$  and recover the FP, we must  lower the concentration
parameter down to $\sim 5$, well below the standard predictions of
numerical simulations of halos in $\Lambda$CDM cosmology (e.g. Wechsler et al., 2002).
\begin{figure*}
\vspace{-1.7truecm}
\centerline{\hspace{4.5truecm}\psfig{file=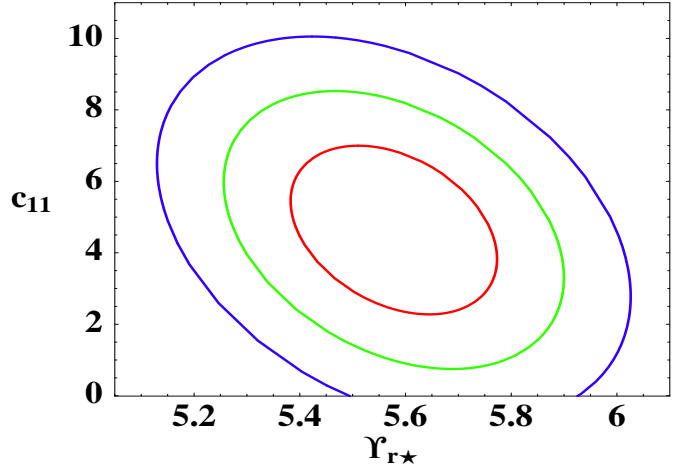,width=200mm,height=250mm}}
\vspace{-16.3truecm}
\caption{Concentration parameter $c_{11}$ and $\Upsilon_{r \ast}$ 
($68\%$, $95\%$ and $99\%$ CL) in agreement with the observed FP, 
for H+NFW mass model with $\Gamma_{vir} = 30$.}
\label{CL30} 
\end{figure*}
Again, it is worth noticing the robustness of the estimate of the
mass--to--light ratio of the TBL component, which results to be insensitive
to even such a model change.

\subsubsection{H+B mass model}

In this case, the presence of dark matter, distributed independently of the stellar
profile, does not alter the FP surface shape, which still remains a plane.
The best--fit mass model is obtained for (see Fig.\ref{FPB}):  
\begin{equation} 
\Upsilon_r = (5.3\pm 0.1)\ \left (\frac{L_r}{L_{\ast}} \right )^{0.21 \pm 0.03}
\label{bestfit1}
\end{equation}
\begin{equation}
\Gamma_e= 0.29 \pm 0.06 
\label{bestfit2}
\end{equation}
(at $68\%$ CL). In Fig.\ref{CLB} ({\it left}) we show the $68\%$, $95\%$ and
$99\%$  confidence contours for the TBL mass--to--light ratio parameters 
$\Upsilon_{r \ast}$ and  $\alpha$.  In the fit,  
the parameter $\Gamma_e$ is somewhat correlated to
$\Upsilon_{r \ast}$: the greater DM amount in the bulge region, the lower the 
mass--to--light ratio of the TBL mass component. 
In Fig.\ref{CLB} ({\it right}) we show  this correlation, marking
the $\Gamma_e$ range corresponding to $99\%$ CL in $\Upsilon_{r \ast}$. 
Notice that a variation of $\Gamma_e$ of a factor $\sim 5$ corresponds to
a much smaller variation of  $\Upsilon_{r \ast}$, giving prominence to
the strong stability  of the stellar mass--to--light ratio we deduce from fits.

Finally, to check the reliability of the fit against our assumption of 
spherical  stellar distribution, we perform the model fit by considering 
only 133 galaxies with small ellipticity ($\epsilon < 0.3$); the resulting 
best--fit parameters are consistent with those of the whole sample,  
with a difference in the mean values of about   $\sim 5 \%$. 

\begin{figure*}
\vspace{-3.5truecm}
\centerline{\hspace{3truecm}\psfig{file=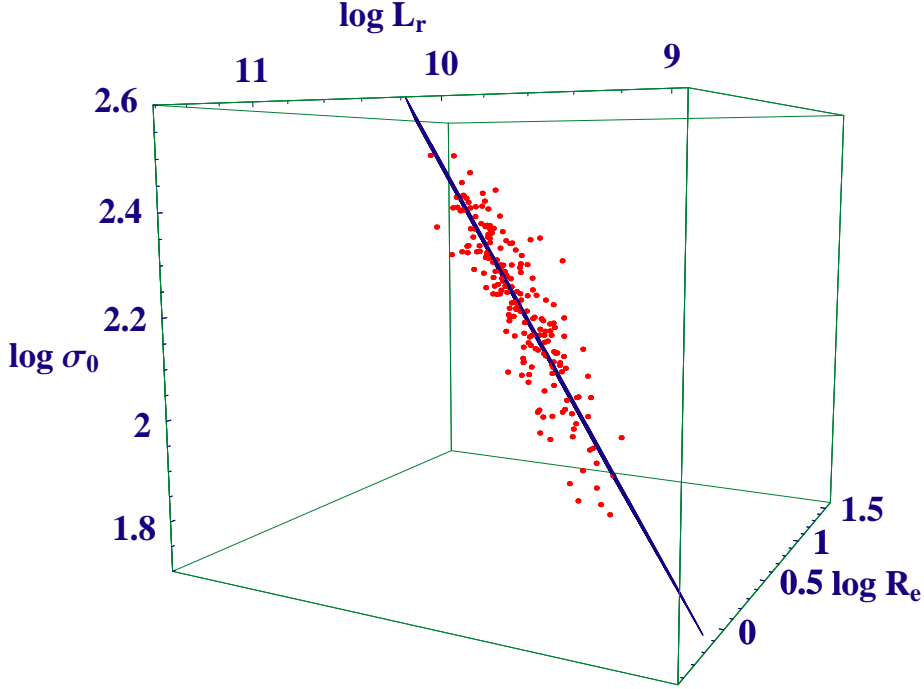,width=230mm,height=330mm}}
\vspace{-19.3truecm}
\caption{The H+B plane (edge--on), best fitting the data. } 
\label{FPB}
\end{figure*}

It is worth stressing that assuming  $\Gamma_e$ constant in the fit is not the most 
general possibility and, in principle, the  DM contribution within $R_e$ 
can well vary with luminosity. Although the investigation of this point is beyond 
the possibilities of our database, we notice  that a variation would be 
in agreement even with recent dynamical and photometric studies 
of mass  distribution in individual ellipticals  (Kronawitter et al., 2000; Gerhard et al., 2001).
Moreover, Gerhard et al. (2001) found $M_{DM}/M_{tot}|_{R_e} \sim 10-40\%$; in comparison we find 
$M_{DM}/M_{tot}|_{R_e} \sim 30\%$.

From eq.(\ref{sigma0B}) we derive the relation between 
$\sigma_0^2 R_e/G$ and the spheroid mass or, equivalently, 
the total mass within the effective radius $M_e \simeq (1+\Gamma_e)\ 0.42\ M_{sph}$:
\begin{equation} 
M_{sph}\  \simeq \ 5.4 \  \frac{\sigma_0^2\ R_e}{G} 
\label{MsphreFP}
\end{equation}
\begin{equation}
M_{e} \equiv M_{sph}(R_e)+M_h(R_e) \simeq \ 2.9 \ \frac{\sigma_0^2\ R_e}{G},  
\label{MereFP}
\end{equation}
using $F_{\rm B}=3.6 \times 10^{-2} $ (see Fig.2) and the best--fit value
$\Gamma_e=0.29$.
Eq.(\ref{MereFP}) explicitly contains the effect of a DM halo and
is to be compared  with the ``gravitational" mass at $R_e$  ($M_G \simeq 2 \ \sigma_0^2\ R_e/ G$),   
found  by Burstein et al. (1997),  taking the standard Keplerian formula 
$M_e=R_e V^2_{rot}/G$ and assuming $V^2_{rot}=3 \sigma^2_0$.
Eq.(\ref{MsphreFP}), instead, is in good agreement with results by 
Ciotti, Lanzoni and Renzini (1996);  indeed they find  for their HP model
(a mass configuration  similar to our H+B model)
$M_{sph} =  c_M \ \frac{\sigma_0^2\ R_e}{G}$ 
with $c_M \simeq 3-6$, according to  the value of the total 
dark--to--TBL mass ratio (in the range $10-70$).

\begin{figure*}
\vspace{0.1truecm}
\centerline{\psfig{file=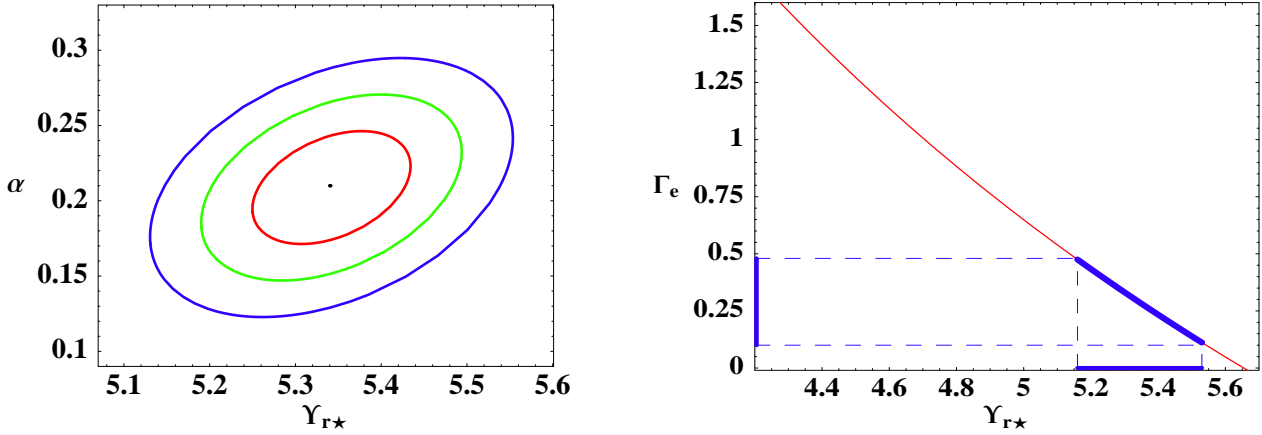,width=190mm,height=290mm}}
\vspace{-22.3truecm} 
\caption{Best--fit H+B mass model: {\it left}) $68\%$, $95\%$ and $99\%$ CL for 
the stellar  mass--to--light parameters;
{\it right}) the correlation between $\Upsilon_{r \ast}$ and $\Gamma_e$, the 
dark--to--stellar mass ratio within $R_e$. Dashed lines 
mark the $99\%$ C.L. in $\Upsilon_{r \ast}$ and $\Gamma_e$.} 
\label{CLB}
\end{figure*} 

In Fig.\ref{ML_L} we show the distribution of the  mass--to--light ratio 
in Gunn--$r$ band of the TBL component, obtained by inserting in eq.(\ref{MsphreFP}) 
the observed $\sigma_0$, $R_e$ and $L_r$. Continuous line is  the  mean 
correlation provided by the FP fit :  
\begin{equation}
\Upsilon_r = 5.3\ (L_r/L_{\ast})^{0.21}
\end{equation}
By  testing  the residuals of the mass--to--light  ratio as function of the effective
radius $R_e$, we find no correlation within the statistical 
errors: $\Upsilon_r \propto R_e^{0.00 \pm 0.05}$.
A possible weak dependence on the effective radius, therefore, seems not sufficient to
justify the scatter observed in the luminosity dependence of $M_{sph}/L$.

Since part of the galaxy sample  has also been  observed in
different photometric bands (JFK, 1992; JFK, 1995a; see Tab.1), we investigated 
the  $M_{sph}/L$ variations with luminosity (for a smaller number of galaxies)
in Johnson $U$ and $B$ and Gunn--$v$ band, obtaining, respectively, the slopes
$0.34\pm 0.05$, $0.27 \pm 0.03$ and $0.25 \pm 0.06$, with similar large scatter,  
thus independent of the photometric band.
The slope of the relation $M_{sph}/L_B \propto L_B^{0.27 \pm 0.03}$   
seems  to be  smaller than  the value of $\sim 0.6 \pm 0.1$, obtained
from  velocity dispersion profiles analysis (Gerhard et al., 2001).
Anyway, this slope tends to higher values when we consider a DM fraction (within
$R_e$) decreasing with galaxy luminosity. For example, we obtain
$M_{sph}/L_B \propto L_B^{0.45 \pm 0.05}$ assuming $\Gamma_e= 1$ for $L_B < 10^{10}
L_{\odot B}$ and  $\Gamma_e= 0.1$ for $L_B > 10^{10} L_{\odot B}$.

\begin{figure*}
\vspace{-2truecm}
\centerline{\psfig{file=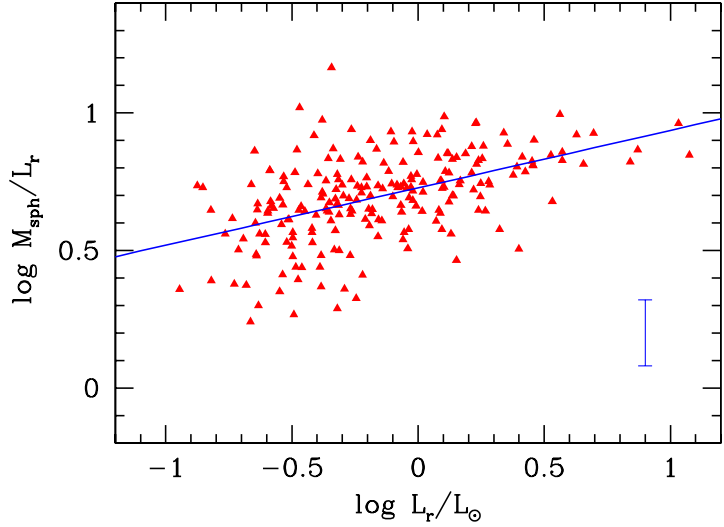,width=120mm,height=120mm}}
\vspace{-1.5truecm}
\caption{The distribution of the stellar mass--to--light ratio in Gunn--$r$ band.
The continous line is  the  mean correlation provided by the FP fit.
The typical data error at 1 $\sigma$ is also shown.}
\label{ML_L}
\end{figure*}

Recent observations have enlightened the potential of galaxy--galaxy (weak) 
lensing to probe the DM halo around galaxies
at large radii, where it has been impossible, so far, to find kinematic tracers. 
Studies of weak lensing by SDSS collaborations (McKay et al., 2002; Guzik and Seljak, 2002)
find $M_{260}/L_r \simeq 110$,  for a $L_{r\ast}$ elliptical galaxy with a NFW dark matter halo 
($M_{260}$ is the mass projected  within an aperture  of radius $260\
h^{-1}$ kpc). Since  the first data bin is at  $R = 75\  h^{-1}$ kpc, the  SDSS g--g lensing 
is  not sensitive to small scales (where NFW and Burkert profiles actually differ);
therefore, we can compare our results with SDSS ones.
We estimate, at large radii, the  dark--to--TBL  mass ratio and the (total) 
mass--to--light ratio of the H+B mass model. In Fig.\ref{massratios}
({\it left}) we show the total (dark+TBL) mass--to--light ratio 
for a $L_{r\ast}$ elliptical galaxy.  Assuming $R_e\simeq 5-10$ kpc for
$L=L_{r\ast}$, the $260\ h^{-1}$ kpc aperture corresponds to 
$\sim (35-70) R_e$, for $h=0.7$.  We obtain the value $M/L_r \simeq 110$ at this 
aperture for  halo core radii in the range $\sim (2.4-2.8)\ R_e$.
This value  for the core  is in agreement with the findings
in spirals (Borriello and salucci, 2001)  and suggests   some form of  DM--baryons interplay as the 
origin of  the soft density cores in halos.

In Fig.\ref{massratios} ({\it right}) we show the cumulative dark--to--luminous 
mass ratio $M_h/M_{sph}$; the value we found above for the halo core radius 
of $r_0 \simeq (2.4-2.8) R_e$  corresponds to a mass ratio, at very large radius, 
of $M_h/M_{sph} \simeq  15-30$. 
Remarkably, this range of values  is in good agreement with results from 
spectro--photometric models of E's formation  in the spheroid 
mass range $3 \times 10^9 \lsim M_{sph} \lsim 2 \times 10^{11}$ (Romano et al., 2002; their Fig.10), 
models including even chimical  evolution and feedback.

\begin{figure*}
\vspace{-0.4truecm}
\centerline{\hspace{1.3truecm}
\psfig{file=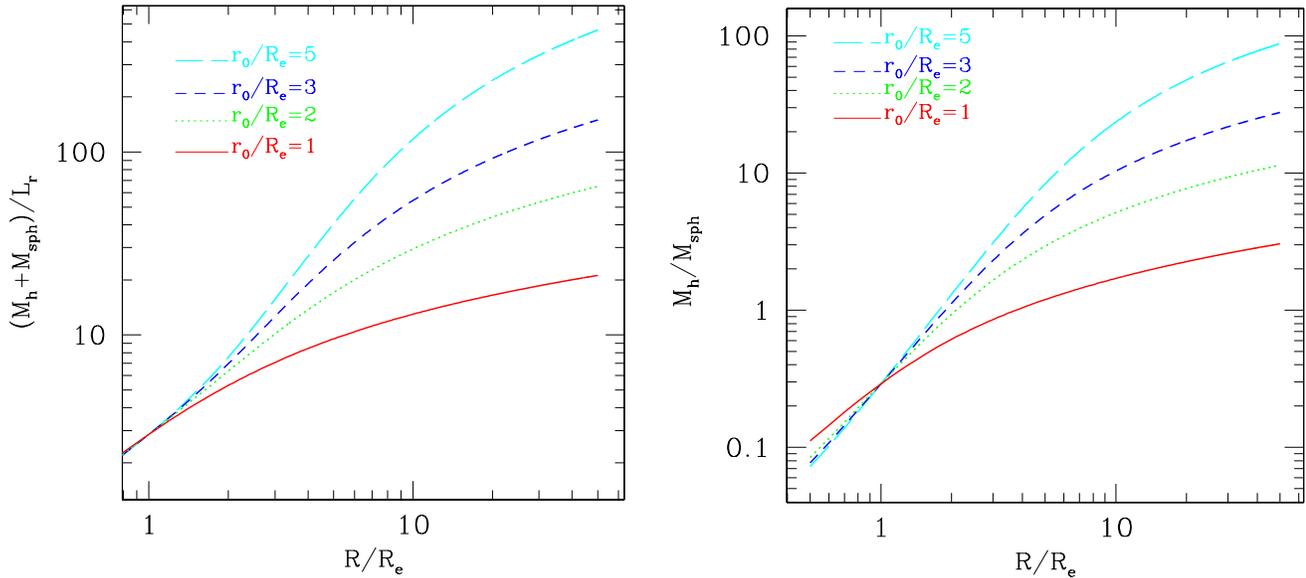,width=230mm,height=320mm}}
\vspace{-22.6truecm}
\caption{Properties of  H+B mass model for a $L_{\ast}$ galaxy
($\Upsilon_r =5.3$, $\Gamma_e=0.3$):
{\it left)}  total (dark+TBL) mass--to--light ratio in Gunn--$r$ band {\it vs.} radius,  
for different values of halo core radius;  {\it right)} dark--to--TBL mass  ratio as function 
of the radius, for different values of $r_0/R_e$.}
\label{massratios}
\end{figure*}

\section{Summary and conclusions}   

We have shown that the very low scatter of elliptical  
galaxies around the Fundamental Plane can be statistically  used to put 
very interesting constraints on DM distribution within them. 
The central velocity dispersion is the key quantity
we have dealt with. We have briefly reviewed its relationship
with the mass distribution of both traced--by--light and dark matter.
Then, we selected a sample of 221 E/S0 galaxies with $L_r\geq 2
\times 10^9$ $L_{\odot}$ in 9 clusters, endowed with very good photometric 
and spectroscopic data. The sample defines the classical FP in the
log--space ($\sigma_0$, $R_e$, $L_r$), with the expected small scatter
(0.084 in $\log\ R_e$, to be compared to a measurement uncertainty
$\Delta\log R_e = \pm 0.045$). 

We tested the reference model of cuspy DM distribution, namely
the NFW model, and the cored model proposed by Burkert (1995).
Our analysis shows that these luminous galaxies
are largely dominated within the effective radius by matter traced by 
light, independently of the DM distribution model, cuspy or cored.
In particular, for the cuspy NFW model, we have shown that
the small scatter of our sample galaxies around the
Fundamental Plane severely challenges the $\Lambda$CDM predictions.
In such a theory, the structural properties of dark and
luminous matter are so interwoven that in the log--space ($\sigma_0$, $R_e$, $L$)
they produce a curved surface, rather than a plane, for plausible values of
the total dark--to--TBL mass ratio. We conclude that
in order to keep the small scatter around the FP we have either to keep
$\Gamma_{vir}$ unacceptably low or to decrease the halo concentration
well below the value currently predicted by simulations in $\Lambda$CDM
cosmology.

Considering a cored DM density distribution,  the agreement 
with the observed FP implies a dark--to--luminous mass fraction within the effective 
radius of  $\sim 30\%$ and a luminosity 
dependence  of the spheroid mass--to--light ratio in Gunn--$r$ band:  
$M_{sph}/L_r=(5.3\pm 0.1) (L_r/L_{\ast r}) ^{0.21\pm 0.03}$.
An important result is the robustness of the mass--to--light ratio of the spheroidal 
component we obtained, which is in good agreement with predictions by stellar 
evolution models.

It is also worth noticing that, besides for spiral and dwarf galaxies,
a cored DM halo, with low internal (within  $2-3 R_e$) density
which increases as $r^{-3}$ at larger radii, is successful to explain also the
structure of {\em elliptical} galaxies, pointing to an intriguing 
homogeneous scenario.
Within this framework, we argue  that dark matter in E's
can be investigated by  a reasonably large number
of galaxies with measures  of  l.o.s. velocity dispersion at  $\sim  R_e$.  
Although, so far, such observations have been
severely hampered by the steep  decreasing of  the surface brigthness with
radius, higher and higher sensitivity  reached by recent surveys offers a good
view  to obtain a better resolution  of the two mass components, in the whole
region  where baryons reside.

\ \

{\bf Acknowledgments}

\noindent The authors would like to aknowledge financial support from  ASI and from   
MIUR  trough COFIN. We also  thank the anonymous referee for very helpful comments.


\newpage

\begin{appendix}

\section{Velocity dispersion in detail}

In this appendix, we will detail the procedure
to compute velocity dispersions by means of Jeans hydrodynamic equations.
Let us set all radii  in units of the effective radius: 
$\r \equiv r/R_e$,  $\R \equiv R/R_e$,  $\rz \equiv r_0/R_e$,
$\rs \equiv r_s/R_e$ and $\rv \equiv r_{vir}/R_e$.

\vspace{0.3truecm}

\noindent {\bf Mass distributions}

\vspace{0.4truecm}

\noindent {\it Traced--by--light mass component} - the radial profiles of mass density, mass 
and surface mass density  are (Hernquist, 1990):

\begin{equation}
\rho_{sph}(\r )=\frac{k}{2\pi}\ \frac{M_{sph}}{R_e^3}\ F_1(\r )
\label{ArhoH}
\end{equation}

\begin{equation}
M_{sph}(\r )=M_{sph}\ F_2(\r )
\label{AMH}
\end{equation}

\begin{equation}
\Sigma(\R )= \frac{1}{2 \pi\ k^2}\ \frac{M_{sph}}{R_e^2}\ F_3(\R )
\label{ASH}
\end{equation}
\noindent where $k \simeq 0.5509$ and:

\begin{equation}
F_1(\r )=\frac{1}{\r \ (\r +k )^3}
\end{equation}

\begin{equation}
F_2(\r )=\frac{\r ^2}{(\r +k)^2}
\end{equation}

\begin{equation}
F_3(\R )=\frac{\left [ \left ( 2+ \frac{\R ^2}{k^2} \right ) X(\R ) -3 \right ]}{(1-\frac{\R 
^2}{k^2})^2}
\end{equation}
\noindent with $X(\R ) = [1-(\R / k)^2]^{-1/2} {\rm Sech}^{-1}(\R /k)$  for $0 \leq \R  < k$  \\ and 
 
$X(\R ) = [(\R / k)^2-1]^{-1/2} {\rm Sec}^{-1}(\R /k)$  for $\R \geq k$.

\vspace{0.4truecm}

\noindent {\it Dark matter halo} - the NFW mass profile reads:

\begin{equation}
M_{NFW}(\r )= M_{sph}\ \Gamma_{vir}\  \frac{A(\r , \rs )}{A(\rv ,\rs )}
\label{AMN}
\end{equation}
\noindent  where, for any pair of variables $(x, y)$,  $A(x, y) \equiv \ln (1+x/y) - x/(x + y)$. 
In particular, recalling that $c \equiv r_{vir}/r_s$,  we have  $A(\rv ,\rs ) = \ln (1+c) - c/(1+c)$.

\noindent The Burkert halo mass profile is:

\begin{equation}
M_B(\r )= 0.416\  M_{sph}\ \Gamma_e\  \frac{B(\r , \rz )}{B(1, \rz )}
\label{AMB}
\end{equation}
\noindent where $B(x, y) \equiv -\arctan (x/y)+2 \ln(1+x/y) + \ln [1+(x/y)^2]$.

\noindent  By inserting eqs. (\ref{ArhoH}), (\ref{AMH}), (\ref{ASH}), (\ref{AMN}) and (\ref{AMB})
in eqs. (\ref{sigmar}), (\ref{sigmap}) and (\ref{sigmaap}) (after variables substitutions),
we will obtain the velocity dispersions profiles:

\newpage

{\noindent \bf Spheroid self--interaction terms}

\noindent Out of the stellar spheroid  self--interaction terms in velocity
dispersions, $\sigma_{r; sph}^2$,  $\sigma_{P; sph}^2$  and $\sigma_{A;
sph}^2$, the first two can be  analitically obtained  (Hernquist, 1990):
\begin{displaymath}
\sigma_{r; sph}^2(\r )= \frac{1}{12\ k}\ \frac{G M_{sph}}{R_e}\ \left [ \frac{12}{k^4} \ \r \ (\r 
+k)\ 
\ln \left ( \frac{\r +k}{\r } \right ) - \frac{\r }{\r +k}\cdot \right .
\end{displaymath}
\begin{equation}
\left . \cdot \left ( 25 + \frac{52}{k}\ \r + \frac{42}{k^2}\ \r ^2 + 
\frac{12}{k^3}\ \r ^3 \right ) \right ]
\end{equation}

\begin{equation}
\sigma_{P; sph}^2(\R )= \frac{1}{6\ k}\ \frac{G M_{sph}}{R_e}\  \frac{F_4(\R )}{F_3(\R )}
\label{AsigpH}
\end{equation}
\noindent where:

\begin{displaymath}
F_4(\R ) =  \frac{1}{2}\ \left (1 - \frac{\R ^2}{k^2} \right )^{-3} 
\left [-3\ \frac{\R ^2}{k^2}\ X(\R )\ \left (8\ \frac{\R ^6}{k^6} 
- 28\ \frac{\R ^4}{k^4} + 35\ \frac{\R ^2}{k^2}  - 20 \right ) + \right .
\end{displaymath}
\begin{equation}
\left . - 24\ \frac{\R ^6}{k^6} +  68\ \frac{\R ^4}{k^4} - 65\ \frac{\R ^2}{k^2} + 6 \right ]  
- 6\ \pi\ \frac{\R }{k} 
\end{equation}
\noindent We obtain the luminosity weigthed  velocity dispersion in the aperture 
$\Ra  \equiv R_A/R_e$, by integrating eq.(\ref{AsigpH}): 

\begin{equation}
\sigma_{A; sph} ^ 2 (\Ra )= \frac{1}{6\ k}\ \frac{G M_{sph}}{R_e}\ 
\frac{ \int_0^{\Ra } F_4(\R  )\ \R  \ d \R  }{ \int_0^{\Ra }  F_3(\R ) \R \  d \R } 
\end{equation}

\noindent where integrals must be numerically performed. For the aperture $\R _a=1/8$, we obtain
the stellar contribution to the   ``central" velocity dispersion:
\begin{eqnarray}
\sigma_{0; sph}^2 & = & \frac{1}{6\ k}\ \frac{G M_{sph}}{R_e}\ \frac{\int_0^{1/8} F_4(\R )\ \R  
\ d\R }{\int_0^{1/8} F_3(\R )\  \R  \ d\R }  \nonumber \\
 & \simeq & 0.174\ \frac{G M_{sph}}{R_e}
\end{eqnarray}

\vspace{0.4truecm}

{\noindent \bf Spheroid--halo  interaction terms}

\noindent We apply the same procedure for calculating the luminous--dark matter
interaction terms: $\sigma_{r; h}^2$, $\sigma_{P; h}^2$ and  $\sigma_{A; h}^2$.
In this case, however, the integrations are always numerical. \\

{\noindent \it H+NFW mass model:}

\begin{equation}
\sigma_{r; NFW}^2(\r )= \frac{G M_{sph}}{R_e}\ \frac{\Gamma_{vir}}{A(\rv , \rs )}\ 
\frac{\int_{\r } ^{\infty} \frac{F_1(\r ')\ A(\r ', \rs )}{\r '^2}\ d\r '}{F_1(\r )}
\end{equation}

\vspace{0.3truecm}

\begin{equation}
\sigma_{P; NFW}^2(\R )= 2\ k^3\  \frac{G M_{sph}}{R_e}\ \frac{\Gamma_{vir}}{A(\rv ,\rs )}\  
\frac{\int_{\R } ^{\infty }   \r \ \frac{\int_{\r }^ {\infty }  \frac{F_1(\r ')\ A(\r ', \rs )}{\r 
'^2}\ d\r '}
{\sqrt{\r ^2-\R  ^2}}\ d \r }{F_3(\R )}
\end{equation}

\vspace{0.3truecm} 

\begin{equation}
\sigma_{A; NFW}^2(\Ra )= 2\ k^3\  \frac{G M_{sph}}{R_e}\ \frac{\Gamma_{vir}}{A(\rv ,\rs )}\  
\frac{\int_0 ^{\Ra }  \R \ \frac{\int_{\R }^{\infty }  \r \ \frac{\int_{\r }^{\infty }  
\frac{F_1(\r ')\ A(\r ', \rs )}{\r '^2}\ d\r '}
{\sqrt{\r ^2-\R  ^2}}\ d \r }{F_3(\R )}\ d\R  }{\int_0^ {\Ra }   F_3(\R )\ \R \ d\R }
\label{AsigaN}
\end{equation}
\noindent where we can reduce the free parameters to the only virial mass {\it i)}
by using eq.(\ref{Rvir}) for the virial radius, with $\Delta_{vir}(z=0)\simeq 337$:

\begin{equation}
r_{vir}  \simeq 2.59 \times 10^{-2}  \left ( \frac{M_{vir}}{M_{\odot}} \right )^{1/3}\ {\rm kpc}
\label{Arvir}
\end{equation}
\noindent and {\it ii)} by assuming  $\Lambda$CDM $c-M_{vir}$ correlation 
(Wechsler et al., 2002, their Fig.16), which, together with eq.(\ref{Arvir}), gives:

\begin{equation}
r_s \equiv \frac{r_{vir}(M_{vir})}{c(M_{vir})} \simeq 6.17 \left ( \frac{M_{vir}}{10^{11}\ M_{\odot}}
\right )^{0.48}\ {\rm kpc}
\end{equation}

\vspace{0.4truecm}

{\noindent \it Burkert halo:}

\begin{equation}
\sigma_{r; B}^2(\r )= 0.416\ \frac{G M_{sph}}{R_e}\ \frac{\Gamma_e}{B(1, \rz )}\  
\frac{\int_{\r }^{\infty } \frac{F_1(\r ')\ B(\r ', \rz )}{\r '^2}\ d\r '}{F_1(\r )}
\end{equation}

\vspace{0.3truecm}

\begin{equation}
\sigma_{P; B}^2(\R )= 0.831\ k^3\  \frac{G M_{sph}}{R_e}\ \frac{\Gamma_e}{B(1,\rz )}\  
\frac{\int_{\R }^ {\infty } \r \ \frac{\int_{\r } ^{\infty } \frac{F_1(\r ')\ B(\r ', \rz )}{\r 
'^2}\ 
d\r '}
{\sqrt{\r ^2-\R  ^2}}\ d \r }{F_3(\R )}
\end{equation}

\vspace{0.3truecm}

\begin{equation}
\sigma_{A; B}^2(\Ra )= 0.831\ k^3\  \frac{G M_{sph}}{R_e}\ \frac{\Gamma_e}{B(1,\rz )}\  
\frac{\int_0^{\Ra } \R \ \frac{\int_{\R }^{\infty } \r \ \frac{\int_{\r } ^{\infty } 
\frac{F_1(\r ')\ B(\r ', \rz )}{\r '^2}\ d\r '}
{\sqrt{\r ^2-\R  ^2}}\ d \r }{F_3(\R )}\ d\R }{\int_0^{\Ra } F_3(\R )\ \R \ d\R }
\label{AsigaB}
\end{equation}

\noindent From eqs. (\ref{AsigaN}) and (\ref{AsigaB}), by setting $\R _a=1/8$, 
we obtain the halos contributions to
the central velocity dispersions given in eq.(\ref{sigma0NFW}) and (\ref{sigma0B}), 
where the functions $F_{NFW}$ and  $F_B$ are defined as:

\begin{displaymath}
F_{NFW}\simeq 2\ \frac{k^3}{\int_0^{1/8} F_3(\R )\ \R \ d\R }\ 
\frac{1}{A(\rv (M_{vir}),\rs (M_{vir}))}\ \cdot 
\end{displaymath}
\begin{equation}
\cdot \int_0^{1/8} \R \ \frac{\int_{\R } ^{\infty } \r \ \frac{\int_{\r } ^{\infty } 
\frac{F_1(\r ')\ A(\r ', \rs (M_{vir} )}{\r '^2}\ d\r '}
{\sqrt{\r ^2-\R  ^2}}\ d \r }{F_3(\R )}\ d\R  
\label{AFNFW}
\end{equation}

\begin{equation}
F_B\simeq 0.83 \  \frac{k^3}{\int_0^{1/8} F_3(\R )\ \R \ d\R }\  \frac{1}{B(1 ,\rz )}\ 
\int_0^{1/8} \R \ \frac{\int_{\R} ^{\infty } \r \ \frac{\int_{\r } ^{\infty } \frac{F_1(\r ')\ B(\r 
', 
\rz )}{\r '^2}\ d\r '}
{\sqrt{\r ^2-\R  ^2}}\ d \r }{F_3(\R )} d\R 
\label{AFB}
\end{equation}
\noindent with the constant  of proportionality: $k^3/ \int_0^{1/8} F_3(\R )\ \R \ d\R \simeq 8.31 
$. 
Let us notice 
that  $F_{NFW}$ depends on both $R_e$ and $M_{vir} 
\equiv \Gamma_{vir}\cdot M_{sph}$, while $F_B$ is only function of the
parameter $\rz \equiv r_0/R_e$.

\end{appendix}

\end{document}